\documentclass[12pt]{article}
\usepackage{epsfig}
\usepackage{amsmath}
\usepackage{amsfonts}

\begin{document}
\pagestyle{empty}
\begin{flushright}
UMN-TH-2535/08\\
\end{flushright}
\vspace*{5mm}

\begin{center}
{\LARGE \bf Dynamical Soft-Wall AdS/QCD}

\vspace{1.0cm}

{\sc Brian Batell}\footnote{E-mail:  batell@physics.umn.edu}
{\small and}
{\sc Tony Gherghetta}\footnote{E-mail:  tgher@physics.umn.edu}
\footnote{Address after January 2008:  School of Physics, The University of Melbourne, Victoria 3010, Australia.}
\\
\vspace{.5cm}
{\it\small {School of Physics and Astronomy\\
University of Minnesota\\
Minneapolis, MN 55455, USA}}\\
\end{center}

\vspace{1cm}
\begin{abstract}
\end{abstract}
We present a solution of the five-dimensional gravity-dilaton-tachyon equations of motion with a pure AdS$_5$ metric and a quadratic dilaton and linear tachyon in conformal coordinates. This leads to an
infrared soft wall model where the dilaton profile gives rise to linear Regge trajectories for the four-dimensional mass spectrum of the dual gauge theory. Even though our approach is phenomenological the scalar fields resemble the dilaton and the closed string tachyon of a non-critical string theory. Interestingly, the linear tachyon has the correct profile to imply that chiral symmetry is not restored for highly excited states. Our solution thus provides a dynamical bottom-up
model of linear confinement in holographic QCD.

\vfill
\begin{flushleft}
\end{flushleft}
\eject
\pagestyle{empty}
\setcounter{page}{1}
\setcounter{footnote}{0}
\pagestyle{plain}

The last decade has seen a renaissance in the study of gauge theories in the nonperturbative regime, spurred by the fascinating idea of gauge/string duality. Inspired by the anti-de Sitter/conformal field theory (AdS/CFT) correspondence \cite{ads1,ads2,ads3}, the basic tenet is that strongly coupled gauge theories can be equivalently described by higher-dimensional gravitational theories. The most obvious application of these ideas is to model Quantum Chromodynamics (QCD), the theory of strong interactions describing the low energy hadrons of our world.
A bottom-up approach to these studies known as AdS/
QCD uses the basic features of QCD to gain an understanding of the hypothetical dual gravitational theory \cite{qcd1,qcd2}.  

In AdS/QCD models, one starts from a weakly coupled five-dimensional (5D) anti-de Sitter background, dual to conformal symmetry of the 4D field theory. This is a reasonable starting point since, due to asymptotic freedom, QCD is approximately conformal at high energies. Bulk fields are added to the 5D theory, which map to QCD operators through the AdS/CFT correspondence. Confinement is implemented either through a hard cutoff in the extra dimension, or by a dynamical cutoff, such as a nontrivial dilaton background \cite{qcddil,qcd3}. The lowest lying meson spectra and decay constants can be modeled quite well with these constructions. However, the main drawback thus far lies in the difficulty of constructing a dynamical model of confinement which produces the correct asymptotic linear Regge trajectories associated with higher radial or spin quantum numbers, as is expected both from the experimental data \cite{pdg} and semiclassical QCD arguments \cite{shifman}.   

In Ref. \cite{linear1}, a heuristic model was proposed in which confinement was modeled with a nontrivial dilaton background, quadratic in the conformal coordinate $z$. This so-called {\it soft wall} correctly produces the linear spectrum in both radial and spin directions. In the simplest model, the mesons follow a mass spectrum $m_{n,S}^2\propto n+S$ for excitation number $n$ and spin $S$ (see also \cite{linear3}). A similar proposal has been put forth which modifies the metric in the infrared (IR) at large $z$ \cite{linear2}.  To date, however, a dynamical model of these setups has not been found. 

In this work, we present a dynamical soft-wall model with a pure AdS metric and quadratic in $z$ ``dilaton'' $\Phi$. Our setup requires an additional scalar field, a ``tachyon'' $T$. Although it is not immediately obvious how to embed our model into string theory, certain features of the scalar potential suggest that $\Phi$ and $T$ may be identified with the dilaton and the closed string tachyon of a non-critical string theory.
At the same time, other aspects of the scalar potential appear rather exotic, as we will discuss further. Nevertheless, our solution is, at a minimum, a phenomenological bottom-up holographic model of linear confinement and dynamical realization of the proposals in \cite{linear1}  and \cite{linear2}, and may yield further insight into the nature of the string theory dual to QCD. We note that an alternative mechanism achieving linear confinement, which relies on open string tachyon condensation, has recently been suggested in \cite{open}.

We will assume that four-dimensional (4D) QCD with a large number of colors $N_c$ can be modeled by the following five-dimensional 
(5D) local action:
\begin{eqnarray}
S&=&M^3\int d^5x \sqrt{-g}\Big[e^{-2\Phi} \Big( R +4g^{MN} \partial_M\Phi \partial_N \Phi\nonumber \\
&&\qquad\qquad\qquad\qquad - \frac{1}{2}g^{MN}\partial_M T \partial_N T -V(\Phi,T) \Big) + e^{-\Phi}{\cal L}_{meson}\Big], 
\label{asf} 
\end{eqnarray}
where $\Phi$ is the dilaton, $T$ is the tachyon, and $M$ is the 5D Planck scale.
One of the points emphasized in \cite{linear1} is that all of the $z^2$ asymptotic IR behavior must be due to the dilaton profile, rather than the metric, to reproduce the linear Regge behavior for high spin mesons. Thus we should look for a solution with a pure AdS metric $g_{MN}=z^{-2}\eta_{MN}$ and $\Phi(z)\sim z^2$. Note that the scalar fields in (\ref{asf}) are dimensionless and the coordinate $z$ in our metric is defined in units where the AdS curvature is unity.

The action (\ref{asf}) is defined in the string frame, but it is simpler to search for solutions in the Einstein frame, which can be obtained with the conformal transformation
\begin{equation}
g_{MN}=e^{4\Phi/3}\widetilde{g}_{MN}.
\label{ct}
\end{equation}
The gravity-dilaton-tachyon action in the Einstein frame is
\begin{equation}
S=M^3\int d^5x \sqrt{-\widetilde{g}} \Big[ \widetilde{R} -\frac{1}{2}\widetilde{g}^{MN} \partial_M\phi \partial_N \phi -\frac{1}{2}\widetilde{g}^{MN}\partial_M T \partial_N T -\widetilde{V}(\phi,T) \Big], 
\label{aef} 
\end{equation}where the tilde distinguishes the two frames, ${\widetilde V} = e^{4\Phi/3} V$, and we have rescaled the dilaton for a canonical action $\phi=\sqrt{8/3}~\Phi$.

We will use the superpotential method \cite{super0,super1} to look for a solution to this system. As an ansatz, we take
\begin{eqnarray}
ds^2&=&e^{-2A(y)}dx^2+dy^2, \label{met2} \\
\phi&=&\phi(y), \\
T &=& T(y).
\end{eqnarray}
The metric (\ref{met2}) is parametrized with the $(x^\mu, y)$ coordinates since the solution techniques are most transparent in these coordinates. The equations of motion become
\begin{eqnarray}
3A''-6 A'^2 &=&\frac{1}{4}\phi'^2+\frac{1}{4} T'^2+\frac{1}{2}\widetilde{V}, \label{eom1}\\
6 A'^2&=&\frac{1}{4}\phi'^2+\frac{1}{4} T'^2-\frac{1}{2} \widetilde{V}, \label{eom2} \\
\phi''-4 A'\phi'& = &\frac{\partial \widetilde{V}}{\partial \phi},  \label{eom3} \\
T''-4 A'T'& = &\frac{\partial \widetilde{V}}{\partial T},  \label{eom4} 
\end{eqnarray}
where $(')$ denotes differentiation with respect to $y$. To proceed we introduce a ``superpotential'' $W(\phi,T)$ to convert the system to a set of first-order equations:
\begin{eqnarray}
A'&=&W, \label{aux1}\\
\phi'&=& 6\frac{\partial W}{\partial \phi}, \label{aux2}\\
T'&=& 6\frac{\partial W}{\partial T}. \label{aux3}
\end{eqnarray}
The potential is then 
\begin{equation}
{\widetilde V}(\phi, T) =18 \Bigg[\left( \frac{\partial W}{\partial \phi} \right)^2+\left( \frac{\partial W}{\partial T}\right)^2\Bigg]-12 W^2.
\label{pot1}
\end{equation}

Our strategy is to specify the metric function $A(y)$ and determine $ \phi, T, W$ and ${\widetilde V}$. 
We will assume a superpotential of the form $W=W_\phi(\phi) + W_T(T)$. Since we would like a quadratic dilaton $\Phi\sim z^2$, the conformal transformation (\ref{ct}) suggests that our metric in the Einstein frame must have the form $\widetilde{g}_{MN}=z^{-2}e^{-2 a z^2}\eta_{MN}$, where $a$ is an arbitrary constant. To apply the superpotential method we must transform to the $y$ coordinates. Explicitly this is given by
\begin{eqnarray}
y&=&\int dz\, \frac{e^{-a z^2}}{z}=\frac{1}{2}{\rm Ei}\left(-a z^2\right), \label{zy2}
\end{eqnarray} 
where ${\rm Ei}(x)$ is the exponential integral function. Now define the inverse function $I$ through the relation
\begin{equation}
z^2=-\frac{1}{a}{\rm Ei}^{(-1)}(2y)\equiv- \frac{1}{a}I(2y).
\label{inverse}
\end{equation}
In fact, we don't need an explicit form for this function; we just need the differentiation rule:
\begin{equation}
\frac{dI}{dy}=\frac{d}{dy}(-a z^2)= - 2 a z^2e^{a z^2}=2I e^{-I}.
\end{equation}
Now using (\ref{inverse}) the function $A(y)$ becomes
\begin{equation}
A(y)=a z^2+\frac{1}{2}\log{z^2}=-I+\frac{1}{2}\log\left(-\frac{I}{a}\right).
\label{ay}
\end{equation}
Differentiating this expression, using (\ref{aux1}), (\ref{aux2}) and (\ref{aux3}), we find
\begin{equation}
 W'= -6Ie^{-2I}+4I^2e^{-2I}=\frac{\phi'^2}{6}+\frac{T'^2}{6}~,
 \label{wprime}
\end{equation}
and can identify
\begin{eqnarray}
\phi'^2&=&24 I^2 e^{-2I}\,,\\
T'^2&=&-36 I e^{-2I}\,.
\label{ep}
\end{eqnarray}
It is straightforward to integrate these equations, which gives rise to two possible solutions for $\phi$ and $T$
\begin{eqnarray}
\phi&=&\pm\,\sqrt{6} I=\mp\,\sqrt{6} a z^2\,, \label{phi} \\
T&=&\pm\, 6 \sqrt{-I}=\pm \,6 \sqrt{a}z\,, \label{T}
\label{e}
\end{eqnarray}
where we have set the integration constants to zero. Remarkably, the dilaton $\phi$ {\it must} have a quadratic profile in order to obtain the desired metric! Remember, we only specified the metric function $A(y)$ in (\ref{ay}), not $\phi$ or $T$. 
The factor $\sqrt{a}$ appearing in (\ref{e}) indicates that $a$ must be strictly positive to avoid a wrong sign kinetic term (ghost) for $T$.
Note that there is the alternative possibility of interchanging the roles of $\phi$ and $T$ from
the ambiguity in the identification in (\ref{wprime}). This would give rise to a linear dilaton and
quadratic tachyon.

Now that we have the background solutions for the scalar fields, it is straightforward to invert these to find the superpotential:
\begin{eqnarray}
W_{\phi}&=&-2\left(1\pm\frac{\phi}{\sqrt{6}}\right)e^{\mp\phi/\sqrt{6}}\,,\\
W_T&=&3 e^{T^2/36}\,.
\end{eqnarray}
These expressions can be substituted into (\ref{pot1}) to find the scalar potential
\begin{equation}
{\widetilde V}_\pm(\phi,T)=\frac{T^2}{2} e^{T^2/18}+ 2\phi^2 e^{\mp2\phi/\sqrt{6}}
-12 \left[3 e^{T^2/36}-2\left(1\pm\frac{\phi}{\sqrt{6}}\right)e^{\mp\phi/\sqrt{6}} \right]^2. 
\label{pot2}
\end{equation}
Reverting back to the string frame (\ref{asf}) and choosing $\phi(z) =\sqrt{8/3}\,z^2$ in (\ref{phi}) with the normalization $a=2/3$ gives the solution
\begin{eqnarray}
\Phi(z)&=&z^2\,,\label{sfPhi}\\
T(z)&=&\pm 2\sqrt{6} z\,,\\
g_{MN}&=&\frac{\eta_{MN}}{z^2}\,,\\
V_-(\Phi,T)&=&e^{-4\Phi/3}{\widetilde V}_-(\phi,T)\,.
\label{sfsoln}
\end{eqnarray}
In this frame the metric is pure AdS$_5$ and the dilaton is quadratic in $z$ realizing the solution posited in \cite{linear1}. Note that the other possible solution $\Phi(z)=-z^2$ gives rise to the metric
$g_{MN}=e^{-8 z^2/3}\eta_{MN}/z^2$, which is only asymptotically AdS$_5$ in the UV limit ($z\rightarrow 0$).

Finally, to make contact with phenomenology, we can introduce the scale $k$ via the coordinate redefinition $z\rightarrow  k z$. For small $z$, the scale $k$ is identified with the AdS curvature, and the bulk cosmological constant is of order $M^3 k^2$. Next we redefine $\phi\rightarrow M^{-3/2}\phi, T \rightarrow M^{-3/2}T$, and $V\rightarrow k^2 V$ in (\ref{asf}). Note that $a$ in our solution is a free parameter, related to an integration constant in (\ref{ay}), and controls the ``position'' of the soft wall $z_0=1/\sqrt{a} k$. Equivalently, the mass scale of the mesons is given by $\sqrt{a}k$.  

The action (\ref{asf}) is reminiscent of the low energy effective action arising from a non-critical string theory, with $T$ identified as the closed string tachyon and the meson action arising from the open string/D-brane sector. In fact, the idea that closed string tachyon condensation in string theory may play a crucial role in screening the gravitational field from the dilaton and maintaining AdS space in the deep infrared was discussed in \cite{linear1,qcd3}. A typical dilaton-tachyon potential (in the Einstein frame) arising from string theory has terms of the form $e^{c\phi}V(T)$, where $c$ is a constant. 
The exponential dilaton factors in (\ref{pot2}) are thus quite similar to what we expect, and notice in particular for the solution with $\Phi=z^2$ there are terms containing $e^{2\phi/\sqrt{6}}=e^{4\Phi/3}$ in the potential. This numerical coefficient in the exponential precisely matches that found in a 5D non-critical string theory \cite{pol1}.

However, the terms linear and quadratic in $\phi$ in (\ref{pot2}) are peculiar. Such terms could perhaps be resummed into the usual exponential form, with additional terms responsible for some modification of QCD dynamics. We might further imagine that the complicated $\phi$ dependence in $\widetilde{V}$, which provides a source for the metric and tachyon backgrounds, could result from integrating out massive states or alternatively could be replaced in a more complete description by fluxes or $\alpha'$ terms in the effective action.

If we identify the dilaton with a string coupling via $g_s=e^{\Phi}$, then for the solution (\ref{sfPhi}) strong coupling occurs in the infrared ($z\rightarrow\infty$). The effective gravity description may still be valid in this regime since the background upon which the strings propagate is pure AdS (i.e. there is no spacetime singularity). 
This is much like the linear dilaton background in which the string frame metric is nonsingular.

Further insight can be gained from the AdS/CFT dictionary, which provides a map between bulk masses and CFT operators. Let us expand the potential to quadratic order in the fields:
\begin{equation}
{\widetilde V}_\pm(\phi,T)\simeq -12-2\phi^2 -\frac{3}{2}T^2+\cdots.
\end{equation}
In the UV limit ($z\rightarrow 0$) the metric is asymptotically AdS, and from the background solutions (\ref{phi}),(\ref{T}) all higher-order terms in the potential are negligible. The first term in the potential is just the negative cosmological constant term. 
We see that the operator dual to $\phi$ (which saturates the Breitenlohner-Freedman bound \cite{bl}) has dimension $\Delta_\phi= 2+\sqrt{4+m_\phi^2}=2$, and does not correspond to any local, gauge-invariant operator in 4D QCD. Although there has been some discussion in recent years of the possible relevance of a dimension two condensate in the form of a gluon mass term \cite{A2}, it is not clear that we should associate $\phi$ with this operator as the AdS/CFT correspondence dictates that bulk fields are dual to gauge-invariant local operators. 
The crucial point is that in order to obtain asymptotically linear trajectories, we only require $\phi\sim z^2 $ in the IR (large $z$). In the UV (small $z$), we should expect the $z$-dependence to change in a more complete holographic description. For example, we might expect $\phi\sim 1 + z^4$ as $z\rightarrow 0$ since the dilaton is proposed to be dual to the dimension four gluon condensate. A simple extension would be to search for a dilaton background which interpolates between the UV behavior dictated by the AdS/CFT dictionary, and the IR quadratic behavior which is necessary for linear confinement. 
More realistically, the running gauge coupling should be incorporated, which requires $\phi \sim - \log \log (-z)$ for small $z$ \cite{qcd3, gk}. Another possible explanation of this strange operator is that our model may be capturing features of supersymmetric QCD, where a dimension two scalar condensate can indeed exist.

The operator dual to the tachyon $T$ has dimension $\Delta_T= 2+\sqrt{4+m_T^2}=3$. It is tempting to associate this bulk field with the QCD operator $\overline{ q} q$. From a string theory perspective, however, it is not clear that this identification is well motivated, since we might expect that the field dual to $\overline{ q} q$ originates from the open string (flavor) sector. 

Nevertheless, a phenomenological approach would be to simply couple ${\cal L}_{meson}$ and $T$ universally to the dilaton and promote $T$ to a charged field. Notice that there is no problem with promoting this field to be charged under the bulk chiral gauge symmetry, since in the action it always appears squared. Interestingly, the linear in $z$ behavior (\ref{T}) is precisely that which is required for the constant mass-squared splittings between chiral partners along Regge trajectories~\cite{open,linearz} and implies the chiral symmetry is not restored in the highly excited states~\cite{csr}. However, the fact that $T$ is linear in $z$ also in the UV indicates that our model corresponds to a hard breaking of the chiral symmetry, or a quark mass, rather than a spontaneous breaking.

From the scalar solutions (\ref{phi}) and (\ref{T}), we see that $\phi \propto T^2$. This raises the possibility that a single scalar field model may exist which exhibits linear confinement. Such a model will likely require a nonstandard kinetic term, or equivalently, higher derivative interactions. We will not investigate this possibility further here, but it would of course be interesting to explore.  

Although beyond the scope of this work, it still remains to check the stability of the model by examining perturbations, in particular those associated with the scalar sector. It is straightforward to show that the tensor and vector modes decouple from the scalar sector, and that zero modes associated with tensor and vector fluctuations are not normalizable. The general perturbation analysis for the case of one bulk scalar field has been performed in \cite{radion}, and we might expect based on these analyses that neither tachyonic nor massless scalar modes exist in our two-field model. These scalar fluctuations are important phenomenologically, being dual to glueball and scalar meson resonances in QCD.

The five-dimensional solution can be generalized to an arbitrary spacetime dimension and with an Einstein frame metric ${\tilde g}_{MN}=z^{-2}\,e^{-2a z^\nu}\eta_{MN}$, where $\nu$ is a real constant. The solution of Einstein's equations in $d+1$ dimensions in the string frame with a pure AdS$_{d+1}$ metric and
$a=2/(d-1)$ is
\begin{eqnarray}
\Phi(z)&=& z^\nu\,, \label{genphi} \\
T(z) &=&\pm4\sqrt{1+1/\nu}\, z^{\nu/2}\,, \label{genT}\\
g_{MN}&=&\frac{\eta_{MN}}{z^2}\,,\\
V_-(\Phi,T)&=&e^{-4\Phi/(d-1)}{\widetilde V}_-(\phi,T)\,,
\end{eqnarray}
where the scalar potential is given by
\begin{equation}
\begin{split}
{\widetilde V}_\pm(\phi,T)
&=\frac{\nu^2}{2}\left(\frac{T^2}{4} e^{T^2/(4(1+1/\nu)(d-1))}+ \phi^2 e^{\mp 2\phi/\sqrt{2(d-1)}}\right)\\
&-d(d-1)\left[(\nu+1) e^{T^2/(8(1+1/\nu)(d-1))}-\nu\left(1\pm\frac{\phi}{\sqrt{2(d-1)}}\right)e^{\mp\phi/\sqrt{2(d-1)}} \right]^2. 
\end{split}
\label{genpot2}
\end{equation}
There is also the solution $\Phi(z)=-z^\nu$ with $g_{MN}=e^{-8z^\nu/(d-1)} \eta_{MN}/z^2$.
In particular it is interesting to analyze the case of 2D QCD (i.e. $\nu=d=2$). As is well known, 't Hooft has demonstrated that 2D QCD in the large $N_c$ limit is exactly solvable and the mesons follow asymptotically linear trajectories~\cite{2Dqcd}. In two dimensions the bulk field $\phi$ corresponds to the dimension two gluon field strength squared $G^2$,  while $T$ is dual to the dimension one $\overline{ q} q$ operator. Our dynamical 3D model thus maps quite well to the 't Hooft model, and it would be interesting to explore this relation further.
For recent work on holographic 2D QCD, see Ref. \cite{2Dqcd2}.

We have presented a phenomenological AdS/QCD construction with a soft infrared wall yielding asymptotic linear Regge trajectories of the high spin mesons. This is the first concrete dynamical model that realizes the proposals of \cite{linear1} and \cite{linear2}. 
At the moment it is not clear how such a model  might be embedded in string theory. It is, however, encouraging that such a confining background indeed exists, and our solution suggests that perhaps closed string tachyon dynamics could play a crucial role in the hypothetical string theory dual to QCD.
\\

\noindent
{\bf Acknowledgements} We thank Andreas Karch, Misha Shifman, Arkady Vainshtein, and Valya Zakharov for useful discussions. We also thank Andreas Karch for comments on the manuscript. This work was supported in part by a Department of Energy grant DE-FG02-94ER40823 at the University of Minnesota, and an award from Research Corporation. T.G. acknowledges the Aspen Center for Physics where part of this work was done.

\end{document}